\documentstyle[prd,aps,preprint,eqsecnum]{revtex}
\begin{document}
\preprint{SUSSEX-AST-95/6-3, IEM-FT-107/95, gr-qc/9506050}
\tighten

\def\half{\frac{1}{2}}
\def\be{\begin{equation}}
\def\ee{\end{equation}}
\def\ba{\begin{eqnarray}}
\def\ea{\end{eqnarray}}

\def\ta{\widetilde{a}}
\def\tg{\widetilde{g}}
\def\tH{\widetilde{H}}
\def\tN{\widetilde{N}}
\def\tp{\widetilde{p}}
\def\tR{\widetilde{R}}
\def\trho{\widetilde{\rho}}
\def\ts{\widetilde{s}}
\def\tT{\widetilde{T}}
\def\tt{\widetilde{t}}
\def\tnabla{\widetilde{\nabla}}

\def\esig{\epsilon_\sigma}
\def\esigast{\epsilon_\sigma^\ast}
\def\nsig{\eta_\sigma}
\def\nsigast{\eta_\sigma^\ast}
\def\epsi{\epsilon_\psi}
\def\epsiast{\epsilon_\psi^\ast}
\def\npsi{\eta_\psi}
\def\npsiast{\eta_\psi^\ast}
\def\dsigma{\delta\sigma}
\def\dpsi{\delta\psi}

\title{Constraints from inflation on scalar--tensor gravity theories}
\author{Juan Garc\'\i a--Bellido and David Wands}
\address{Astronomy Centre, School of Mathematical and
Physical Sciences, \\
University of Sussex, Brighton BN1 9QH.~~~U.~K.}
\date{\today}
\maketitle
\begin{abstract}

We show how observations of the perturbation spectra produced during
inflation may be used to constrain the parameters of general
scalar-tensor theories of gravity, which include both an inflaton and
dilaton field.  An interesting feature of these models is the
possibility that the curvature perturbations on super-horizon scales
may not be constant due to non-adiabatic perturbations of the two
fields. Within a given model, the tilt and relative amplitude of the
scalar and tensor perturbation spectra gives constraints on the
parameters of the gravity theory, which may be comparable with those
from primordial nucleosynthesis and post-Newtonian experiments.

\end{abstract}


\section{Introduction}

The most convincing explanation for the flatness, isotropy and
homogeneity of the observed universe is the inflationary scenario
\cite{Book}.
Moreover, the most compelling evidence for this model is the prediction
of a nearly scale-invariant distribution of Gaussian perturbations.
If these are indeed the origin of the perturbations observed in the
microwave background sky, and of the initial inhomogeneities from which
galaxies formed, then they could provide our earliest glimpse of the
physics of the early universe and, in particular, of the effective
theory of gravity at that time.

In this paper, we consider the possible constraints that can be
placed upon the allowed gravity theory during inflation.
Precision tests of gravity in the solar system severely constrain the
effective gravity theory today~\cite{Will93}, while
predictions from primordial nucleosynthesis have been used to restrict
scalar-tensor
deviations from general relativity since the radiation dominated
era~\cite{Kalligas}. Our aim is to push back those limits to a
still earlier epoch, i.e. inflation.

We will do this within the context of general scalar-tensor gravity
theories, which involve a massless dilaton or Brans-Dicke
field~\cite{Brans+Dicke61,STT}. These provide a well-defined class of
theories against which to test the predictions of general
relativity. Almost all attempts to produce a renormalizable quantum
theory of gravity seem to include scalar fields non-minimally coupled
to the spacetime curvature.  The low-energy effective action of string
theory, for instance, involves a dilaton field coupled to the Ricci
curvature \cite{string}.  Scalar fields coupled directly to the
curvature tensor appear in all dimensionally reduced gravity theories,
and their influence on cosmological models was first seriously
considered by Jordan~\cite{Jordan59}.  Gravity Lagrangians including
terms of higher order in the Ricci scalar can also be cast as
scalar-tensor theories\cite{Tey+Tou83,Wands94} with appropriate scalar
potentials.

Recently, Damour and Nordtvedt \cite{Dam+Nor93} have pointed out that
during a dust dominated cosmological era, scalar-tensor theories of
gravity tend to approach general relativity at late times even in the
absence of a potential for the Brans-Dicke field.  This can be
parametrized by the Brans-Dicke parameter $\omega$, which determines
the ratio of the scalar to tensor couplings to matter, and tends to
infinity in the general relativistic limit.  Damour and Nordtvedt
realized that this would occur during most of the history of the
universe, when the trace of the energy-momentum tensor drives the
scalar field. However, this mechanism is not effective during the
radiation dominated era~\cite{RADSOLNS}. Considering a wide class of
theories where $\omega$ is some arbitrary function of the
gravitational coupling, Damour and Nordtvedt~\cite{Dam+Nor93}
calculate how far towards the general relativistic limit the universe
might be expected to evolve (i.e.~how large $\omega$ becomes) after
the radiation era.  They find that it should be simply related to the
number of e-foldings, $N=\ln(a_0/a_{\rm eq})$ since matter-radiation
equality.

As most of the expansion of the universe to date occurred during the
inflationary era, one expects the same approach to general relativity
to occur during inflation~\cite{HyperInf90}. This has been recently
discussed in the context of string theory by Damour and
Vilenkin~\cite{Dam+Vil95}. Thus, apart from setting the scene for the
conventional hot big bang, by producing a spatially flat, isotropic,
homogeneous (but slightly perturbed) metric, it is tempting to suggest
that inflation may also produce general relativity as the low-energy
effective theory of gravity, even if it started as a generic
scalar-tensor theory at the Planck scale~\cite{JGBDW}.

We show in this paper that this can indeed occur. We give the
generalization to scalar-tensor gravity of the inflationary slow-roll
parameters in general relativity and show that the vanishing of these
parameters corresponds to the usual post-Newtonian limit of Einstein
gravity. We calculate the scalar and tensor perturbations produced
during inflation in a general scalar-tensor theory and evaluate the
relative amplitude and slope of the spectra in terms of the slow-roll
parameters. While it is difficult to obtain model-independent bounds
on the scalar-tensor parameters, we will study a particular model of
chaotic inflation for which strong bounds can be given that are
comparable with the post-Newtonian and nucleosynthesis limits.

\section{Scalar--tensor gravity theories}
\label{sectstt}

The scalar--tensor field equations are derived from the action~\cite{STT}
\begin{equation}
S = \int d^4x \sqrt{-\bar g} \left[ f(\phi) \bar R
	- \half \bar g^{ab} \phi_{,a} \phi_{,b} - U(\phi)
	+ \bar{\cal L}_{\rm matter} \right] \; ,
\label{eSTTaction}
\end{equation}
where $\bar R$ is the usual Ricci curvature scalar
and $16\pi f(\phi)$ is the Brans--Dicke field.
Thus the gravitational coupling strength (Newton's constant $G$
in general relativity) is determined here by the dynamical
variable $f(\phi)$. In the particular case of Brans-Dicke gravity,
$\ f(\phi)=\phi^2/8\omega$, where $\omega$ is the Brans-Dicke
parameter, and we recover general relativity in the limit $\omega \to
\infty$. More general scalar-tensor theories with different choices
of $f(\phi)$ correspond to the case where $\omega(f)$ is a function of $f$.

The potential $U(\phi)$ is the generalization of the cosmological
constant $\Lambda$ in general relativity. Perturbation spectra
produced in models of inflation driven by a potential for the
Brans-Dicke field have recently been discussed by
Kaiser~\cite{Kaiser}, while the bounds on the allowed mass have been
discussed by Damour and Vilenkin~\cite{Dam+Vil95} and Steinhardt and
Will~\cite{Ste+Wil94}. Such a potential is often introduced to fix the
value of Brans-Dicke field at late times, however we shall show that
this is not necessary in order to attain general relativity as a
cosmological attractor. In what follows we will leave $f(\phi)$ as a
free function but consider only models in which $U(\phi)$ is
zero. They correspond to flat directions in the scalar potential of
string effective theories \cite{Casas}.  In the absence of
non-perturbative effects, the dilaton remains massless. We will work
in this approximation and will discuss possible extensions in future
work.

Matter is minimally coupled to the metric $\bar{g}_{ab}$ and thus test
particles follow geodesics in this frame, which we refer to as the
Jordan frame. However it is often useful to write the action in terms
of the conformally related Einstein metric~\cite{Dicke}
\be
g_{ab} \equiv e^{-2a(\psi)}\,\bar{g}_{ab}\ ,
\ee
where the conformal factor is defined as $e^{-2a}\equiv2\kappa^2f(\phi)$,
in terms of which the action in Eq.~(\ref{eSTTaction}) takes the
Einstein-Hilbert form with a fixed gravitational constant
$\ G\equiv\kappa^2/8\pi$, and the Brans-Dicke field can be written as
a scalar field $\psi$ with a canonical kinetic term in the new matter
lagrangian
\be
{\cal L}_{\rm matter} = -\half g^{ab}\psi_{,a}\psi_{,b} + e^{4a(\psi)}
\bar{\cal L}_{\rm matter} \ .
\ee
There are now explicit interactions between this field and the original
matter fields whose energy-momenta are therefore not necessarily
conserved with respect to $g_{ab}$ \cite{Dicke,CQG}.
We could define a dimensionless parameter~\cite{Dam+Nor93},
$\ \alpha(\psi) \equiv \kappa^{-1}(da/d\psi)$,
which characterizes the scalar-tensor theory: $\alpha^2$ specifies the
ratio of the dilaton and graviton couplings to matter. A given choice
of function $a(\psi)$, or equivalently $f(\phi)$, determines
$\alpha(\psi)$. In particular, it is related to the
Brans-Dicke parameter $\omega$ by
\be
2\alpha^2 = {1\over2\omega+3} \ .
\label{alpha}
\ee
For a linear $a(\psi) = \alpha\kappa\psi$ with constant $\alpha$, we
recover Brans-Dicke theory. \footnote
{Our notation coincides with that of Damour-Gibbons-Gundlach~\cite{DGG} and
Starobinsky-Yokoyama~\cite{Sta+Yok95} for constant $-2\alpha = \gamma
= \beta/2$.}

Present day observational tests constrain the Post-Newtonian parameters
$\gamma$ and $\beta$ \cite{Will93}, written in terms of
$\alpha$ and $\alpha'(\psi) \equiv \kappa^{-1}(d\alpha/d\psi)$ as
\be\label{PPN}
\gamma = 1 - {4\alpha^2\over1 + 2\alpha^2} \ , \hspace{6mm}
\beta = 1 + {2\alpha^2\alpha'(\psi)\over1 + 2\alpha^2}\ ,
\ee
which are constrained by present-day solar system tests to be
\cite{Will93} $\ |\gamma - 1| < 0.002\ $ and
$\ |4\beta - \gamma - 3| < 0.001\ $ \cite{NMG7}.
Therefore,
\be\label{PNB}
\alpha^2 < 5 \cdot 10^{-4}\ , \hspace{6mm}
4\alpha^2 |1 + 2\alpha'| < 10^{-3} \ .
\ee
There are similar constraints from primordial
nucleosynthesis~\cite{Kalligas}. These constrain possible variations
of the Planck mass during and after the radiation dominated era.  Our
aim is to go beyond the radiation era and try to constrain possible
deviations from general relativity during inflation.

\section{Scalar-tensor inflation}

In this section we will analyze the classical evolution of the scalar
fields during inflation. The inflaton field, $\sigma$,
minimally coupled in the Jordan frame, with a self-interaction
potential $V(\sigma)$ gives an explicit matter lagrangian
to consider in a scalar-tensor cosmology. The total matter Lagrangian
in the Einstein frame including the Brans-Dicke field is then
\be
{\cal L}_{\rm matter} =
 - \half g^{ab} \psi_{,a} \psi_{,b}
 - \half e^{2a(\psi)}\, g^{ab} \sigma_{,a} \sigma_{,b}
 - e^{4a(\psi)}\, V(\sigma) \ .
\ee

We see by inspecting the potential term in the Lagrangian  that $\sigma$
will evolve towards a minimum of $V(\sigma)$ while  $\psi$ evolves
towards a minimum of $a(\psi)$ or, equivalently, a zero of
$\alpha(\psi)$. However, from Eq.~(\ref{alpha})  we see that a zero of
$\alpha(\psi)$ requires that $\omega\to\infty$. That is, general
relativity will generically be an attractor during the
cosmological evolution, if $a(\psi)$ possesses a minimum.

The field equations for the fields $\sigma$ and $\psi$
in a spatially flat Friedmann-Robertson-Walker metric are then
\ba
\ddot\sigma + 3H\dot\sigma &=& -\ 2 \alpha \kappa\,\dot\psi\,
\dot\sigma - e^{2a}\,V'(\sigma)\ , \\[2mm]
\ddot\psi + 3H\dot\psi &=& -\ \alpha \kappa\Big( e^{4a}\,4V
- e^{2a}\,\dot\sigma^2 \Big)\ , \\[2mm]
\dot H &=& -\ {\kappa^2\over2} \left( \dot\psi^2
       + e^{2a}\,\dot{\sigma}^2 \right)\ ,
\label{eHdot}
\ea
and the Hamiltonian constraint,
\be
H^2 = {\kappa^2 \over 6} \left(\dot\psi^2 + e^{2a}\,\dot\sigma^2 +
e^{4a}\,2V\right) \ .
\label{econ}
\ee

The condition for inflation to occur in the Einstein frame
$\ |\dot H| < H^2\ $ is thus, see Eqs.~(\ref{eHdot}) and (\ref{econ}),
\be
\dot\psi^2 + e^{2a}\,\dot\sigma^2 < e^{4a}\,4V(\sigma)\ .
\ee

\subsection{Slow-roll inflation}

We will work in the slow-roll approximation in both scalar fields.
In principle this is not a necessary constraint: one of the fields
might roll quickly to the minimum of its potential and then the problem
reduces to single field inflation, either the familiar chaotic
inflation in general relativity (for $\dot\psi = 0$) or old extended
inflation in Brans-Dicke (for $\dot\sigma = 0$). However, we would like
to consider the more general case in which both fields
slow-roll~\cite{Mol+Mat92,GBLL,JDB}.
In this case, the general field equations can be written as first-order
equations,
\ba
3H^2 &\simeq& \ \kappa^2 e^{4a}\,V(\sigma) \ ,\\[2mm]
3H \dot\sigma &\simeq& -\ e^{2a}\,V'(\sigma) \ ,\\[2mm]
3H \dot\psi &\simeq& -\ 4\alpha\kappa\,e^{4a}\,V(\sigma) \ .
\label{SLOW}
\ea

Neglecting the other terms in the equations of motion amounts to the
following assumptions
\ba
{\rm max}\left\{\dot\psi^2, \ \alpha^2 \dot\psi^2, \ e^{2a}\,\dot\sigma^2
\right\} \ll e^{4a}\,V(\sigma) \ ,\\[2mm]
|\ddot\sigma| \ll |H \dot\sigma| \qquad {\rm and} \qquad
|\ddot\psi| \ll |H \dot\psi| \ .
\ea

Having written down first-order equations for the evolution of
the fields we can turn the slow-roll assumptions based on values of the
fields' derivatives into consistency equations in terms of the
parameters of the theory:
\ba
\esig \equiv {1\over2\kappa^2} \left({V'(\sigma)\over V(\sigma)}
\right)^2 \ , &\hspace{2cm}&
\nsig \equiv {1\over\kappa^2}\,{V''(\sigma)\over V(\sigma)}\ , \\[3mm]
\epsi \equiv 8 \alpha^2(\psi)\ , &\hspace{2cm}&
\npsi \equiv 4 \alpha'(\psi) - 16 \alpha^2(\psi)\ .
\ea
The consistency equations for slow-roll inflation are then $\left\{
e^{-2a}\,\esig,\, e^{-2a}\,|\nsig|,\, \epsi,\, |\npsi|\right\} \ll 1$.
The first two conditions are just the expected generalization to
scalar-tensor gravity of the slow-roll conditions for an inflaton
field in general relativity. Notice, however, that if $\ \dot{a}
= \kappa\alpha \dot\psi \leq 0\ $ during the subsequent evolution
of the universe (i.e.~$a>0$), the conditions on $\esig$ and $\nsig$
are relaxed compared to the general relativistic case, where
$a(\psi)=0$ throughout. The last two parameters are not present in
general relativity and arise here by requiring that both the
Brans-Dicke and the inflaton field slow-roll.

We have defined the slow-roll parameters $\epsilon$ and $\eta$ by
extending the usual definition of these parameters for a single field
\cite{LidParBar94} to the separable potential for the two fields
$U(\sigma,\psi)=V(\sigma)\,e^{4a(\psi)}$. It is intriguing to note that
the limit of vanishing slow-roll parameters for the $\psi$ field
coincides with the general relativistic weak-field limit in the
post-Newtonian parametrization of the scalar-tensor gravity
theory \cite{Will93}, see Eq.~(\ref{PPN}).

In calculating the rate of change of quantities with respect to
different co-moving scales, it is useful to write down the
relation between the number of $e$-foldings from the end of
inflation and the values of the scalar fields,
\be
\label{Nefold}
N = - \int_{t_e} Hdt\
 = {\kappa\over4}\,\int_{\psi_e} \,{d\psi\over\alpha(\psi)} \
 = \,\kappa^2\,\int_{\sigma_e} {V(\sigma)\over
e^{2a}V'(\sigma)}\,d\sigma \ .
\ee
Our present horizon crossed outside the Hubble scale about 50-60 e-foldings
before the end of inflation. In fact, the precise number depends
logarithmically on the energy scale during inflation and
the efficiency of reheating, and so is weakly model-dependent.

\section{Density perturbations}

Inflation is the only known mechanism that solves the horizon and
homogeneity problems. However, the main observational constraint on
inflationary models is the spectrum of density perturbations that they
produce. Strictly speaking, observations of perturbations in the
microwave background, or of the large-scale structure in our patch of
the universe, only provides an upper limit on the level of density
perturbations, which could perhaps be produced by some other source of
inhomogeneities. Nonetheless, the apparently Gaussian and nearly
scale-invariant nature of the perturbations are natural properties of
perturbations due to quantum fluctuations of the inflaton field during
inflation.

In the case of a single slow-rolling field, only adiabatic
perturbations are possible. Any fluctuation in the field must produce
a fluctuation in the local curvature. However, in the presence of two
coupled fields we must also consider the effect of isocurvature (or
entropy) perturbations between the two fields. In parti\-cular this
can lead to the breakdown of the usual assumption that the curvature
perturbation is frozen-in on scales greater than the Hubble length
$(k_{\rm ph} < H)$.  It is important to allow for such effects if we
hope to constrain the possible role of a variable gravitational
coupling.

Our calculations extend those of Starobinsky and Yokoyama
\cite{Sta+Yok95} who considered the parti\-cular case of inflation in
Brans-Dicke gravity. As we shall see, their results are readily
extended to more general scalar-tensor theories.

\subsection{Perturbed field equations}

In this section we will consider the linear perturbations around the
homogeneous background fields, $\sigma(t) + \dsigma(t,x^i)$, $\psi(t)
+ \dpsi(t,x^i)$, with a perturbed metric
\be
d s^2 = - \left(1+2\Phi(t,x^i)\right)dt^2
 + R^2(t)\left(1-2\Phi(t,x^i)\right) \delta_{ij}dx^idx^j \ .
\ee
We can study the evolution of each Fourier mode (whose physical
wavenumbers we denote by $k_{\rm ph}$) separately, since they decouple
in the linear approximation.  The perturbed field equations then yield
the following expressions to first order
\ba
\ddot{\dsigma} + 3H\dot{\dsigma} + k_{\rm ph}^2\dsigma
&+& e^{2a}\,V''(\sigma)\dsigma = 4\dot\sigma\dot\Phi +
- 2 e^{2a}\,V'(\sigma)\Phi
\nonumber \\[3mm]
-\ 2\alpha\kappa\Big(\dot\sigma\dot{\dpsi} &+& \dot\psi\dot{\dsigma}\Big)
- 2\alpha'(\psi)\kappa^2\,\dot\sigma\dot\psi\,\dpsi
- 2\alpha\kappa\,e^{2a}\,V'(\sigma)\dpsi \ ,\label{PPSI}
\ea
\ba
\ddot{\dpsi} + 3H\dot{\dpsi} + k_{\rm ph}^2\dpsi
&+& \alpha'(\psi)\kappa^2\left(e^{4a}\,4V - e^{2a}\,\dot\sigma^2
\right)\dpsi + 2\alpha^2\kappa^2\left(e^{4a}\,8V - e^{2a}\,\dot\sigma^2
\right)\dpsi \nonumber \\[3mm] = 4\dot\psi\dot\Phi
&-& 8\alpha\kappa e^{4a}\,V \Phi -
2\alpha\kappa\left(e^{4a}\,2V'(\sigma)\dsigma - e^{2a}\,
\dot\sigma\dot{\dsigma}\right) \ ,\label{PSIG}
\ea
\ba
\ddot\Phi & + & 4H\dot\Phi + \left(\dot H + 3H^2\right)\Phi \nonumber \\
& = & {\kappa^2\over2}
\left[\dot\psi\dot{\dpsi} + e^{2a}\,\dot\sigma\dot{\dsigma}
- e^{4a}\,V'(\sigma)\dsigma + \left(e^{2a}\,\dot\sigma^2
- e^{4a}\,4V\right) \alpha\kappa\,\dpsi \right] \ ,\label{PPHI}
\ea
together with the energy and momentum constraints
\ba
\dot\Phi + H\Phi & = & {\kappa^2\over2} \left(\dot\psi\dpsi + e^{2a}
\dot\sigma\dsigma\right) \ ,\label{DPHI} \\[3mm]
3H\dot\Phi & + & \left(\dot H + 3H^2\right)\Phi + k_{\rm ph}^2\Phi
 \nonumber \\
& = & -\ {\kappa^2\over2}
\left[\dot\psi\dot{\dpsi} + e^{2a}\,\dot\sigma\dot{\dsigma}
+ e^{4a}\,V'(\sigma)\dsigma + \left(e^{2a}\,\dot\sigma^2
+ e^{4a}\,4V\right) \alpha\kappa\,\dpsi \right] \ .\label{EMP}
\ea

A very useful quantity for the study of perturbation spectra is the
three-curvature of comoving hypersurfaces~\cite{BST83,Lyth85},
\be
\label{DEFZETA}
\zeta \equiv \,
-\, {H^2\over\dot H}\,\left( \Phi + H^{-1}\dot\Phi \right) + \Phi \ .
\ee
Combining Eqs.~(\ref{PPHI})--(\ref{EMP}), and using the equations of
motion, we find an exact expression for the time variation of $\zeta$,
\be\label{DZETA}
\dot\zeta = \, k_{\rm ph}^2\, {H\over\dot{H}}\,\Phi -\,
H \left[ {d\over dt} \left( {e^{2a}\dot\sigma^2 - \dot\psi^2
 \over e^{2a}\dot\sigma^2 + \dot\psi^2} \right)
 + \dot{C} \right] \left({\dpsi\over\dot\psi} -
{\dsigma\over\dot\sigma}\right) \ ,
\ee
where $\dot{C}=2\alpha\kappa\dot\psi e^{4a}
\dot\sigma^4/(e^{2a}\dot\sigma^2+\dot\psi^2)^2$ is due to the
frictional damping of the $\sigma$ field by $\psi$.

In the single field case (where one of the fields is held fixed) the
right-hand side of Eq.~(\ref{DZETA}) vanishes in the limit
$k_{\rm ph} \to 0$, and thus $\zeta$ remains constant on scales exceeding
the Hubble length~\cite{Lyth85}. This allows one to determine the
large-scale curvature perturbation at the end of inflation simply by
equating it with the perturbation when the mode first crossed outside
the Hubble scale. However this is true in general only for adiabatic
perturbations and need no longer hold in the presence of two
fields~\cite{DGL,Sta+Yok95}.

This is due to the entropy perturbation~\cite{MukFelBra92}
\be
\tau \delta S = \,\dot H
\left[ {d\over dt} \left( {e^{2a}\dot\sigma^2 - \dot\psi^2
 \over e^{2a}\dot\sigma^2 + \dot\psi^2} \right)
 + \dot{C} \right]
\left({\dpsi\over\dot\psi} - {\dsigma\over\dot\sigma}\right)
\ .
\ee
The first term in the square brackets will be present whenever two
fields are evolving but the second term, $\dot{C}$, would not be
present if both fields had standard kinetic terms.  It is clear that
for adiabatic modes $\dpsi/\dot\psi =\dsigma/\dot\sigma$
(perturbations along the classical trajectory) $\zeta$ remains
constant on super-Hubble scales, but in general the curvature
perturbation at horizon crossing cannot be equated with that at the
end of inflation, due to the non-adiabatic terms.

Note that all the above perturbed equations are exact and we have not
yet invoked the slow-roll approximation. In the following we shall
solve for the evolution of $\zeta$ to lowest order in the slow-roll
parameters.

We will assume that the curvature perturbation will be conserved on
super-Hubble scales through re-heating and the subsequent radiation
and matter dominated eras. Recently, it has been emphasized~\cite{Der+Muk95}
that the curvature perturbation $\zeta$ is conserved at
transitions in the equation of state across a boundary at a fixed
energy density. This is the case when perturbations are adiabatic and
the end of inflation must coincide with a particular energy density.
However this is not necessarily the case in the presence of two
fields. We can see from Eq.~(\ref{DZETA}), that our assumption that
only adiabatic perturbations exist at the end of inflation requires
that the motion of one of the fields dominates. This will usually be
the case, especially when inflation ends because one of the fields'
kinetic energy becomes comparable to the potential energy. The
slow-roll parameter $\epsilon$ for one of the fields becomes of order
unity while the other remains small. It is possible that both
slow-roll parameters become large at the same point, but this seems to
be unlikely a priori.

The perturbation at the end of inflation can then be equated with that at
re-entry if its subsequent evolution remains adiabatic. Any variation
of the Brans-Dicke field after the end of inflation could invalidate
this assumption. Nucleosynthesis limits~\cite{Kalligas} suggest this
does not occur at temperatures below about an MeV and in the absence
of a potential for the Brans-Dicke field, $\dot\psi=0$ is a stable
solution during a radiation dominated era for arbitrary scalar-tensor
gravity theories~\cite{RADSOLNS}. However this could be
altered by the presence of an explicit potential term and the
consequences would require careful investigation.

\subsection{Short-wavelength limit}

For large values of $k_{\rm ph}\gg H$ we can neglect the potential
terms in the perturbed field equations~(\ref{PSIG}) and~(\ref{PPSI})
and they reduce to those for massless fields (i.e.,
$e^{-2a}|\eta_\psi|$, $|\eta_\sigma|\ \ll1$).  Thus, to lowest order
in the slow-roll parameters, the expectation values of the
perturbations as they cross outside the Hubble radius ($k_{\rm
ph}\simeq H_\ast$) are given by Gaussian random variables with $\
\langle|\dsigma_\ast|^2\rangle = e^{-2a_\ast}H_\ast^2/2k^3 \ , \ \
\langle|\dpsi_\ast|^2\rangle = H_\ast^2/2k^3 \, ,\ $
where $k$ is the comoving wavenumber. Note that, while the field
$\psi$ is minimally coupled in the Einstein frame, the $\sigma$ field
is minimally coupled in the Jordan frame and therefore it is the conformally
transformed Hubble constant (to lowest order) that determines its
amplitude at Hubble crossing~\cite{JGBNuclPhys}.

We shall denote the spectrum of a quantity $A$ by $\ {\cal P}_A (k)
\equiv {4\pi k^3 \over (2\pi)^3}\ \langle|A|^2 \rangle\, ,\ $
as defined in~\cite{Lid+Lyt93}. Thus we have
\ba
{\cal P}_{\dsigma}
 \simeq e^{-2a_\ast}\, \left( {H_\ast \over 2\pi} \right)^2 \ , \\
{\cal P}_{\dpsi}
 \simeq \left( {H_\ast \over 2\pi} \right)^2 \ .
\ea


\subsection{Long-wavelength limit}

For slowly varying ($\dot\Phi \ll H\Phi$), long-wavelength
($k_{\rm ph} \to 0$) modes, to lowest order in the slow-roll
parameters, the Eqs.~(\ref{DPHI}), (\ref{PPSI}) and (\ref{PSIG})
reduce to
\ba
\Phi &\simeq& -\ 2 \alpha\,\kappa\,\dpsi - \half\,{V'(\sigma)\over V}\,
\dsigma \ ,
\\[2mm]
3H\dot{\dpsi} &\simeq& \ 4\alpha'(\psi)\,\kappa^2\,e^{4a}\,V\,\dpsi \ ,
\\[2mm]
3H\dot{\dsigma} &\simeq& -\ e^{2a}\,\left(V'(\sigma)\over
V\right)'\,V\,\dsigma
+ 2\alpha\,\kappa\,e^{2a}\,V'(\sigma)\,\dpsi \ .
\label{DDSIG}
\ea
Note that for constant $\alpha$ we recover Starobinsky and Yokoyama's
results \cite{Sta+Yok95}.

Using Eqs.~(\ref{SLOW}), the last two equations
can be integrated to give the evolution of
fluctuations in the scalar fields at long-wavelengths:
\ba
\dpsi &\simeq& - {4\alpha\over\kappa} Q_1 \ , \label{defdpsi} \\
\dsigma &\simeq& {1\over\kappa^2}{V'(\sigma)\over V}
 \left( Q_2 -  e^{-2a} Q_1 \right) \ ,
\label{defdsig}
\ea
and thus
\be
\Phi \simeq 8 \alpha^2 Q_1 - {1\over2\kappa^2} \left( {V'(\sigma)\over V}
\right)^2 \left( Q_2 - e^{-2a} Q_1 \right) = \epsi Q_1 - \esig
 \left( Q_2 - e^{-2a} Q_1 \right) \ ,
\label{defPhi}
\ee
where $Q_1$ and $Q_2$ are constants of integration, chosen to coincide
with those introduced by Starobinsky and Yokoyama \cite{Sta+Yok95}.
It will be convenient to define a new constant $Q_3\equiv
Q_1 e^{-2a_\ast} - Q_2$, so that $Q_1$ and
$Q_3$ are independent Gaussian random variables whose values, for a
given Fourier mode, are determined by the amplitude of $\dsigma_*$ and
$\dpsi_*$ at horizon-crossing (when $k_{\rm ph} = H_\ast$).
Thus they have expectation values
\ba
{\cal P}_{Q_1} = {e^{4a_\ast}\,\kappa^4\,V_\ast \over 24\pi^2\,
\epsi^\ast} \label{expQ1} \ , \\
{\cal P}_{Q_3} = {e^{2a_\ast}\,\kappa^4\,V_\ast \over 24\pi^2
\,\esig^\ast} \label{expQ3} \ .
\ea

{}From Eqs.~(\ref{DPHI}) and~(\ref{DEFZETA}) we have, during slow-roll
\cite{Mol+Mat92,GBLL}, in the long wavelength limit,
\be
\zeta \simeq H\,{\dot\psi\dpsi + e^{2a}\dot\sigma\dsigma\over
\dot\psi^2 + e^{2a}\dot\sigma^2} \ .
\ee
As shown in Eq.~(\ref{DZETA}) this expression need not be constant.
Substituting in our results for the long-wavelength modes of the
scalar fields, we have
\be
\label{SRZETA}
\zeta \simeq {\left( \epsi + (e^{-2a}-e^{-2a_*})\esig\right) Q_1 +
\esig Q_3 \over \epsi + e^{-2a} \esig} \ .
\ee
If either of the scalar fields is fixed ($\esig$ or $\epsi$
identically zero) then we recover the single field results where
$\zeta$ is constant (equal to $Q_1$ or $Q_3$ respectively).

The spectrum of density perturbations at the end of inflation
${\cal P}_{\zeta_e}(k)$ can be computed from (\ref{SRZETA}),
\be
\label{PKZETA}
{\cal P}_{\zeta_e} \simeq
\left({\epsi^e + (1 - e^{-2a_*})\esig^e\over\epsi^e + \esig^e}
\right)^2 {\cal P}_{Q_1} + \left({\esig^e\over\epsi^e +
\esig^e}\right)^2 {\cal P}_{Q_3}\ .
\ee

In Sect.~\ref{NumSect} we will study a particular model and give numerical
results showing how and when the different terms dominate.

\subsection{Gravitational wave perturbations}

In addition to the scalar curvature perturbations that give rise to
density perturbations, tensor or gravitational wave perturbations can
also be generated from quantum fluctuations during
inflation~\cite{Abbott+Wise83}. Since we have chosen to work in the
Einstein conformal frame, we can use the standard results for the
evolution of tensor perturbations of the metric. The two independent
polarizations evolve like minimally coupled massless fields with a
spectrum~\cite{MukFelBra92,Lid+Lyt93}
\be
{\cal P}_g = {2\kappa^4 e^{4a_\ast}V_\ast \over 3\pi^2} \ .
\ee

Graviational wave perturbations can contribute to the microwave
background anisotropies only on the largest scales (scales larger than
the Hubble scale at last-scattering, corresponding to about $>1^\circ$
on the sky). Their contribution relative to scalar curvature
perturbations is given by the ratio~\cite{Lid+Lyt93}
\be
\label{RATIO}
R \simeq {3\over4} {{\cal P}_g \over {\cal P}_{\zeta_e}} \ .
\ee
The rapid decay of the gravitational wave anisotropies on smaller
scales is their most distinctive signature. In Sect.~\ref{NumSect} we
will study a particular model and show how the ratio $R$ changes with
scale.

\section{Observational constraints}

Having allowed for the possible evolution of the curvature
perturbation $\zeta$ on super-Hubble scales during inflation, we will
now restrict our analysis to those cases where $\zeta$ has become
constant on observable scales by the end of inflation, i.e.,
entropy pertubations become negligible. This allows us
to assume that $\zeta$ remains fixed on super-Hubble scales until it
re-enters the Hubble length at late times. We can then relate the
curvature perturbation at the end of inflation to the density
perturbation at re-entry during the matter dominated era,
\be
\label{dH}
\delta_H^2 \simeq {4\over25} {\cal P}_{\zeta_e}\ ,
\ee
following the notation of~\cite{Lid+Lyt93}.

In any model of inflation, the amplitude of the density perturbations
depends upon the magnitude of the potential energy density relative to
the Planck scale, which is essentially a free parameter.  We will
concentrate upon the variation of the amplitude of the curvature
perturbations with co-moving scale. At each point in the spectrum, the
`tilt' is given by the spectral index $n_s$, where $\ n_s - 1 =
d\ln\delta_H^2/d\ln k\ $.  This can be evaluated within the slow-roll
approximation, where the comoving wavenumber corresponds to a given
scale at horizon crossing, $d\ln k\simeq -dN_*$, and thus from
Eq.~(\ref{Nefold}),
\be
\label{TILT}
n_s - 1 = {d\ln\delta_H^2 \over d\ln k} \simeq
 - \left( {4\alpha_\ast\over\kappa} {\partial\over\partial\psi_*}
  + {e^{2a_\ast}V_\ast'\over\kappa^2V_\ast}
    {\partial\over\partial\sigma_\ast} \right) \ln\delta_H^2\ .
\ee
A well-known result in general relativity, for slow-roll inflation
with a single field, is that $n_s-1 \simeq - 6\epsilon_\ast +
2\eta_\ast$, a function solely of the slow-roll parameters at
Hubble-crossing~\cite{tilt,Lid+Lyt93}.  Because $\zeta$ can evolve on
super-Hubble scales during scalar-tensor inflation, $\zeta_e$ and thus
$\delta_H^2$ will in general depend upon parameters both at Hubble
crossing and at the end of inflation. However, the latter will not
change with co-moving scale.

Large-angle microwave background experiments probe scales close to our
observable horizon, which crossed outside the Hubble scale when
$N\sim60$~\cite{Lid+Lyt93}. The COBE two-year data constrains the
spectral tilt to be in the range $0.7<n<1.7$ at the $1\sigma$
level~\cite{Gorski94}, but future experiments should be able to
constrain the tilt to within about $0.1$~\cite{CMBtilt}. However, from
observations of galaxy clustering we might hope to recover the
primordial density perturbations on scales down to about 1Mpc, which
leave the Hubble scale at about $N\sim50$. This could provide a much
more precise determination of the tilt, though it is currently limited
by uncertainties in other cosmological parameters such as the value of
the Hubble constant or the type of dark matter. While a tilt of
$0.7<n<0.8$ may have some advantages over $n=1$ in an otherwise
standard cold dark matter model, a tilt $n<0.6$ is probably
unacceptable~\cite{tiltvalue,Lid+Lyt93}.

The general expression for the tilt from Eq.~(\ref{TILT}) is rather
complicated. We will only attempt to evaluate it in two general
regions of parameter space and then specialize it to the case
of an inflaton with a generic chaotic inflationary potential, for
which we give numerical results.

Note that the tilt of the gravitational wave spectrum is just given by
\be
n_g \simeq - 16 \alpha_\ast^2 - 2 e^{2a_\ast} \esig^\ast \ .
\ee
Unlike the approximate expressions for the scalar tilt which will be
given below, this simple expression for $n_g$ is valid in the whole
range of parameter space. Moreover, since both terms on the
right-hand-side must be non-positive, one could in principle give a
direct constraint on $\alpha^2$ completely independent of the form of
the inflaton potential. However, the measurement of this slope will be
exceedingly difficult. Tensor perturbations do not contribute to
structure formation and in many inflationary models the observable
effect of gravitational waves is completely negligible~\cite{Lid+Lyt93}.

The main constraint coming from gravitational waves will be their
relative amplitude, given by Eq.~(\ref{RATIO}). If $R$ becomes of
order unity then, since independent Gaussian random variables add in
quadrature, the amplitude of the scalar perturbations inferred from
anisotropies of the microwave background on large scales is reduced by
about $70\%$. As $R$ increases, the allowed amplitude of scalar
perturbations decreases, eventually becoming incompatible with
structure formation. It is the combined effect of a tilted spectrum
and the gravitational wave contribution that proves such a strong
constraint on models of extended inflation~\cite{tilt}.

Finally, note that if $n_s<1$ then, on very large scales, the
potential energy density relative to the Planck mass at Hubble
crossing becomes so large that the amplitude of density perturbations
is of order unity and the universe enters a self-reproducing regime,
where the classical motion is dominated by quantum
fluctuations~\cite{GBLL,JGBNuclPhys,JGBDW}. Such inhomogeneities, even
on super-horizon scales today, could be detected through the
Grishchuk-Zel'dovich effect~\cite{GZ} unless they are on scales at
least 500 times greater than our horizon~\cite{GZ500}. This represents
another constraint on any model. Sufficient inflation in the classical
regime thus requires $N_{\rm max}>66$.

\subsection{Scalar-tensor extended inflation}

Let us first consider the case where the Brans-Dicke field evolution
dominates that of the inflaton at the end of inflation,
$\epsi^e\gg\esig^e$. Then, for scales crossing outside the Hubble scale
near the end of inflation, we have $\zeta_e\simeq Q_1$ and thus $\
\delta_H^2 \simeq (4/25) {\cal P}_{Q_1}\ $.  This remains valid
at scales for which $(\epsi^e/\esig^e)^2 \gg
e^{-2a_\ast}\epsi^\ast/\esig^\ast$. It includes models of extended
inflation~\cite{ExtInf} where the field $\sigma$ is trapped in a
metastable false vacuum so that $\esig=0$ (for any $\nsig>0$) and
where $\zeta$ remains fixed on super-Hubble scales. But this result
for $\delta_H$ also includes perturbations for which
$\esig^\ast\geq\epsi^\ast$, where there will be significant evolution
of the curvature perturbation $\zeta$ when $k_{\rm ph}<H$ during
inflation.

We then have
\be\label{OLDTILT}
n_s - 1 \simeq - 16 \alpha_\ast^2 + 8 \alpha_\ast'
 - 2e^{2a_\ast}\esig^\ast
 \ .
\ee
When $\esig^\ast=0$ this expression generalizes the well known result
for extended inflation in Brans-Dicke models~\cite{Lid+Lyt93} to more
general scalar-tensor theories. We see that, just as in general
relativity, $n_s$ need not always be less than unity. For instance, we
can produce a Harrison-Zel'dovich spectrum ($n_s=1$) by choosing a
scalar-tensor theory where $\alpha'=2\alpha^2$ corresponding to
$a(\psi)=-(1/2)\ln(\psi/\psi_e)$. This is a particular realization of
``intermediate inflation''~\cite{Bar+Lid93}.

More generally, as $\esig$ is always non-negative, a lower bound on the
tilt of the power spectrum then constrains the slow-roll parameters of
the gravity theory, irrespective of the form of the inflaton potential.

The relative contribution of tensor and scalar perturbations to the
microwave background anisotropies is given by Eq.~(\ref{RATIO}), which in
this limit yields
\be
\label{OLDR}
R \simeq 10^2\, \alpha_\ast^2 \ .
\ee
An upper limit on this ratio then constrains $\alpha^2$ independently of
$\alpha'$.

\subsection{Scalar-tensor chaotic inflation}

In the opposite limit, $\esig^e\gg\epsi^e$, in which the evolution of
the inflaton dominates that of the Brans-Dicke field at the end of
inflation, we find $\zeta_e\simeq Q_3$ and thus $\ \delta_H^2
\simeq (4/25) {\cal P}_{Q_3}\ $. This result will hold for the
last scales to leave the Hubble length during inflation. It remains
valid on larger scales as long as
\be
\label{Q3GREATER}
\left(1-e^{-2a_\ast}+{\epsi^e\over\esig^e}\right)^2 \ll
{\epsi^\ast\over e^{2a_\ast}\esig^\ast}
\ee
is satisfied, see Eq.~(\ref{PKZETA}).
In these limits, the curvature perturbation is always due to
fluctuations in the inflaton field $\sigma$, but there may still be
evolution on super-Hubble scales due to the frictional damping by
$\psi$. We find from Eq.~(\ref{SRZETA}) that $\zeta\simeq e^{2a}Q_3$,
which coincides with the solution of $\dot\zeta\simeq\dot{C}\zeta$, given by
Eq.~(\ref{DZETA}) in this limit.

Given the above result for $\delta_H^2$ we thus find that on
sufficiently small scales the tilt will be given by
\be
\label{nSLOWROLL}
n_s -1 \simeq e^{2a_\ast}\left(-6\esig^\ast+2\nsig^\ast\right)
 - 8\alpha_\ast^2 \ .
\ee
Note that $\nsig$ can be positive or negative and thus could lead to a
positive spectral tilt~\cite{Lid+Lyt93,BlueSpectra}.

The larger effective gravitational constant at early times
($a_\ast>0$) amplifies the tilt due to the changing shape of the
inflaton potential, and its variation leads to an additional negative
tilt. Any chance of constraining $\alpha_\ast^2$ from observations of
the tilt is clearly limited by uncertainty in the form of the inflaton
potential. In the simplest case of chaotic inflation driven by a
polynomial potential, $V(\sigma)=\lambda\sigma^{2n}/2n$, there is a
simple relation between $\esig$, $\nsig$ and the value of $\sigma$,
\be
\esig = \left({n \over 2n-1}\right) \nsig
 = {2n^2 \over \kappa^2\sigma^2} \ .
\ee
This guarantees that $-6\esig+2\nsig$ is negative and thus the slope
of the density perturbation spectrum in Eq.~(\ref{nSLOWROLL}) is
always $n_s<1$. In arbitrary scalar-tensor theories, a lower limit on
the slope then places an upper bound on $\alpha^2$.

The ratio between the scalar and tensor contributions to the microwave
background anisotropies reduces to the usual general relativistic
case~\cite{Lid+Lyt93},
\be
\label{RSLOWROLL}
R \simeq 12\, e^{2a_\ast}\,\esig^\ast \ .
\ee

When the condition given in Eq.~(\ref{Q3GREATER}) no longer holds and
instead $(1-e^{-2a_\ast})\gg\epsi^\ast/\esig^\ast$, we find that
$\zeta_e\simeq Q_1$. In this regime the results of Eqs.~(\ref{OLDTILT})
and~(\ref{OLDR}) apply.
It is interesting to note that the naive calculation based on taking
$\zeta\simeq\zeta_\ast$ does in fact give the correct result and, even
in our careful analysis, $\zeta$ remains constant on super-Hubble
scales. This is clearly seen in the Fig.~1, where we
show the evolution of $\zeta$ after Hubble crossing
in the specific model of chaotic inflation discussed in the next
section.

This occurs despite the fact that at Hubble crossing the curvature
perturbation is due to the $\psi$ field, $\zeta\simeq
H\delta\psi/\dot\psi$, while by the end of inflation it appears as a
perturbation in the $\sigma$ field, $\zeta\simeq
H\delta\sigma/\dot\sigma$. This is a consequence of the coupling
between the two fields and the dependence of $\delta\sigma$ upon the
evolution of $\delta\psi$ seen in Eq.~(\ref{DDSIG}).  We do not expect
this result to hold in general for two fields in general
relativity. Moreover, for intermediate scales $\zeta$ does evolve on
super-Hubble scales.

\section{Numerical results}
\label{NumSect}

In this section we will try to show the main features discussed above
with a particular example that includes both regimes.
We will choose the arbitrary origin of the field $\psi$ so that
$\psi_e=0$ at the end of inflation (when $a_e=0$) and assume that
$\alpha(\psi)$ can be approximated as a linear
function during the latter stages of inflation (when observable scales
cross outside the horizon), i.e., we take a Taylor expansion for
$a(\psi)$ up to second order, as was done in~\cite{Kalligas}.
\ba
a(\psi) = a_1\,\kappa\psi
 + { a_2\over2}\left(\kappa\psi\right)^2 \ ,\\
\alpha(\psi) = a_1 +  a_2\,\kappa\psi \ ,
\ea
where $a_2$ and $a_1$ are constants and $\psi$ is then given
in terms of the number of $e$-foldings, using Eq.~(\ref{Nefold}), as
\be
\kappa\psi \simeq {a_1\over a_2}
\left[ \exp\left(4 a_2\,N\right) - 1 \right] \ ,
\ee
Therefore we have
\be
\alpha \simeq a_1 \exp\left(4 a_2\,N\right) \ ,
\ee
and
\be
e^{-2a} = 2\kappa^2 f(\phi) \simeq \exp\left[-\,{a_1^2\over a_2}\,
\left(\exp(8 a_2\,N) - 1\right)\right] \ .
\ee
Note that at the end of inflation ($N=0$) we have the present value of
the gravitational coupling, and our parameters $a_1$ and $a_2$
correspond to $\alpha$ and $\alpha'$ respectively, at the end of
inflation.

For a simple inflaton potential
$V(\sigma) = \lambda\sigma^4/4$, the slow-roll solution for $\sigma$,
satisfying $\esig^e = 1$, is
\be
\kappa^2\sigma^2 \simeq 8 + {\exp(a_1^2/ a_2)\over a_2}\,
\left[E_1(\alpha^2/ a_2) - E_1(a_1^2/ a_2)\right]\ ,
\ee
where $E_1(z)$ is the exponential integral function \cite{AS}.

It will be convenient to define some new variables
\be\label{new}
x = \kappa\sigma \ , \hspace{2cm} y = \exp(-2\,a(\psi))\ ,
\ee
whose classical evolution in the slow-roll approximation is shown in
Fig.~2 for $ a_2=10^{-2}$ and $a_1=10^{-2}$ and
$a_1=10^{-3}$.  Note that in both cases $y$, and thus the Planck
mass, becomes essentially constant by the end of inflation.

If we also introduce $w\equiv1-y+8a_1^2$, then in the limit $w_\ast^2
\ll y_\ast x_\ast^2\alpha_\ast^2$, where all starred quantities are to be
evaluated at $N=N_\ast$, we satisfy the condition given in
Eq.~(\ref{Q3GREATER}) and we find $\zeta_e\simeq Q_3$. This must hold
for the last modes to leave the Hubble scale at the end of inflation,
as $y\to1$ and $x\to2\sqrt{2}$, for $8a_1^2\ll1$.  For larger scales,
when $w_\ast^2 \gg y_\ast x_\ast^2\alpha_\ast^2$, we find
$\zeta_e\simeq Q_1$. This is clearly demonstrated in Figs.~3(a)
and~3(b).

The spectral tilt (\ref{TILT}) at a scale which crosses outside the
Hubble length when $N=N_*$, is given, at lowest order in the slow-roll
parameters, by
\be
\label{TAST}
n_s - 1 \simeq
 - 8 \left[ { \left(2\alpha_\ast^2 -  a_2 +(2/y_\ast x^2_\ast)\right)
    w_\ast^2
 + 2\alpha_\ast^2 y_\ast w_\ast
 + \alpha_\ast^2 \left(3 + y_\ast x^2_\ast\alpha_\ast^2\right)
\over w_\ast^2 + y_\ast x_\ast^2\alpha_\ast^2} \right] \ ,
\ee
see Fig.~4. In Fig.~5 we have also
computed the ratio of gravitational to scalar components,
\be
R \simeq {96\alpha_*^2 (8a_1^2 + 1)^2
\over w_*^2 + y_* x^2_*\alpha^2_*}\ .
\ee
We thus recover the results of Eqs.~(\ref{nSLOWROLL})
and~(\ref{RSLOWROLL}) for $n_s-1$ and $R$ respectively on small
scales, and on larger scales by Eqs.~(\ref{OLDTILT})
and~(\ref{OLDR}). The variation of both the tilt and the ratio $R$
with Hubble crossing epoch, $N_\ast$, is shown in Figures~4 and~5.

Observational constraints on $R$ and the tilt of the perturbation
spectrum can bound the values of the parameters $a_1$ and
$a_2$, thus constraining deviations from general relativity as far
back as $N_\ast$ $e$-folds before the end of inflation. Figure~6 is a
contour plot showing $n_s$ for scales that left the horizon at $N_\ast=60$,
corresponding roughly to $6000$ Mpc today, and thus the sort
of scale constrained by observations of large-scale structure. We see
that both $a_1$ and $ a_2$ must be very small in order for the
tilt of the spectrum to remain close to the general relativistic value
of $n_s\simeq0.95$. This reflects the need to keep the Planck mass
essentially constant to avoid large departures from the
Harrison-Zel'dovich ($n_s=1$) spectrum. Figure~7 shows similar results
for the contribution of gravitational wave perturbations at the
same scale, $N_\ast = 60$.

\section{Conclusions}

In this paper we have considered the constraints that may be placed
upon the effective theory of gravity during a period of inflation in
the early universe. We do this in the context of scalar-tensor
theories, taking the coupling of the Brans-Dicke field to
matter as an arbitrary function $\alpha(\psi)$, and neglecting any
explicit potential for the dilaton field.

Present day observational limits on the variation of the Brans-Dicke
field are expressed as bounds on the post-Newtonian parameters of the
theory. We have shown that the general relativistic limit of these
parameters coincides with the vanishing of the corresponding slow-roll
parameters for the Brans-Dicke field during inflation. Slow-roll
inflation already requires the scalar-tensor theory to be close to the
general relativistic limit. The observed spectrum of density
perturbations produced from quantum fluctuations in the inflaton and
Brans-Dicke fields can then constrain just how large the deviation may
be.

A careful calculation of the curvature perturbation $\zeta$ during
inflation shows that some of the results, applicable to single-field
inflation in general relativity, no longer apply. Due to the evolution
of the two fields during inflation, there will be non-adiabatic
perturbations, which can lead to the evolution of $\zeta$ on scales
larger than the Hubble length. Therefore, the amplitude of $\zeta$ at
re-entry can no longer be equated with that at the time the scale left
the horizon. This is a general feature of inflation with two
fields. However, in scalar-tensor theories there are regimes for which
$\zeta$ does remain constant outside the Hubble scale. In models where
inflation ends as the inflaton field rolls to the minimum of its
potential, we find two regimes for which $\zeta$ at the end of
inflation is equal to that at Hubble-crossing. It is only in an
intermediate regime that the naive calculation breaks down.

We give expressions for the spectral slopes, $n_s$ and $n_g$, and
relative amplitude, $R$, of the scalar and tensor perturbations
produced, in terms of the slow-roll parameters of a general
scalar-tensor theory of gravity. Observational bounds then place
constraints on these parameters. A possible signature of scalar-tensor
inflation is the breakdown of the consistency relations predicted in
single-field inflation~\cite{recon}.

It is important to emphasize that our ability to make quantitative
predictions relies on our knowledge of the inflaton potential.  In our
specific example of a chaotic inflation model, we find that if we
constrain the slope of the spectrum to be $n_s>0.6$ we obtain bounds on
$\alpha$ and $\alpha'$ at the end of inflation that are comparable
with those from nucleosynthesis or solar-system tests. For
example, for $\alpha<0.015$ we require $\alpha'<0.01$, which is
much stronger than the corresponding post-Newtonian bound.

Future observations will be able to constrain $n_s$ to within
$0.1$~\cite{CMBtilt} which would further improve the bounds on
($\alpha$,$\alpha'$), in the context of a given inflaton
potential. Although far harder to measure, the tilt of the tensor
perturbations, $n_g$, gives a model-independent bound on $\alpha$.

\section*{Acknowledgements}

JGB and DW acknowledge support from PPARC.  The authors are grateful
to Andrew Liddle and Jos\'e Mimoso for useful discussions. DW
acknowledges use of the Starlink computer system at Sussex.



\section*{Figure Captions}

\noindent
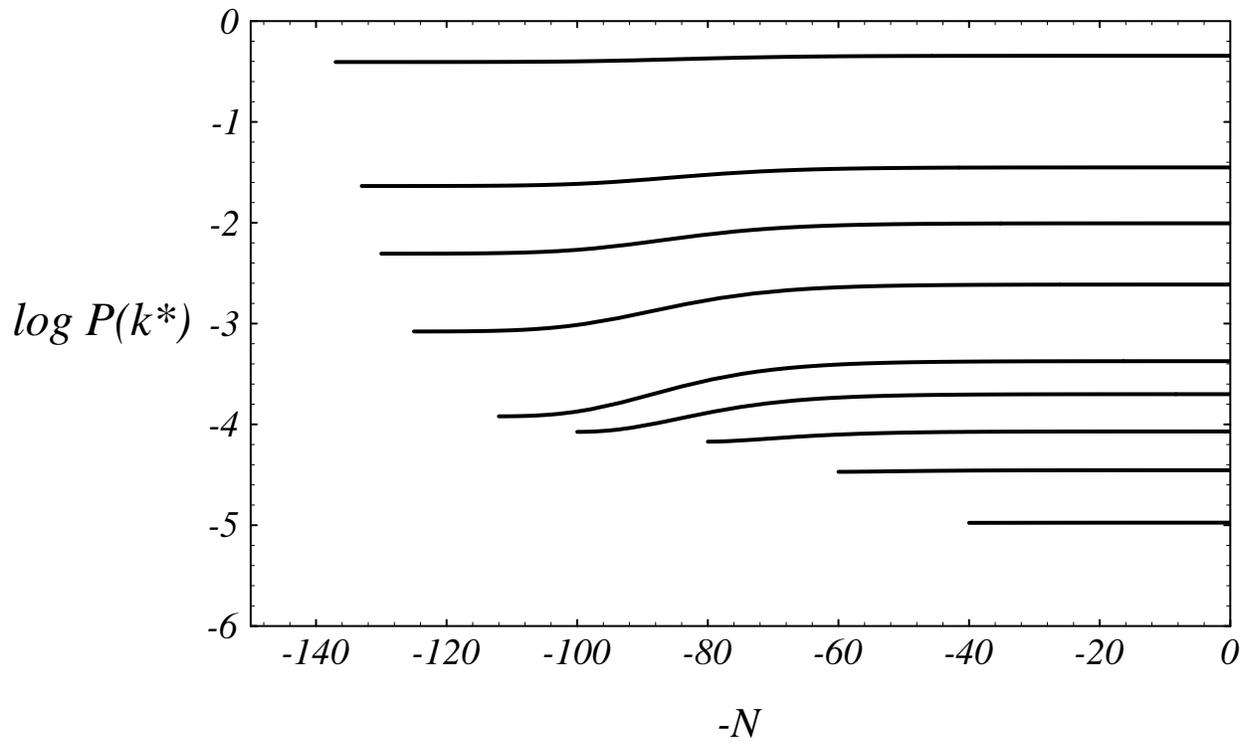
Fig.~1. The evolution of the amplitude of density perturbations after
Hubble-crossing for a range of comoving scales, $k_*$, as a function
of the number $N$ of $e$-folds from end of inflation, for the
model described in Sect.~\ref{NumSect}.

\

\noindent
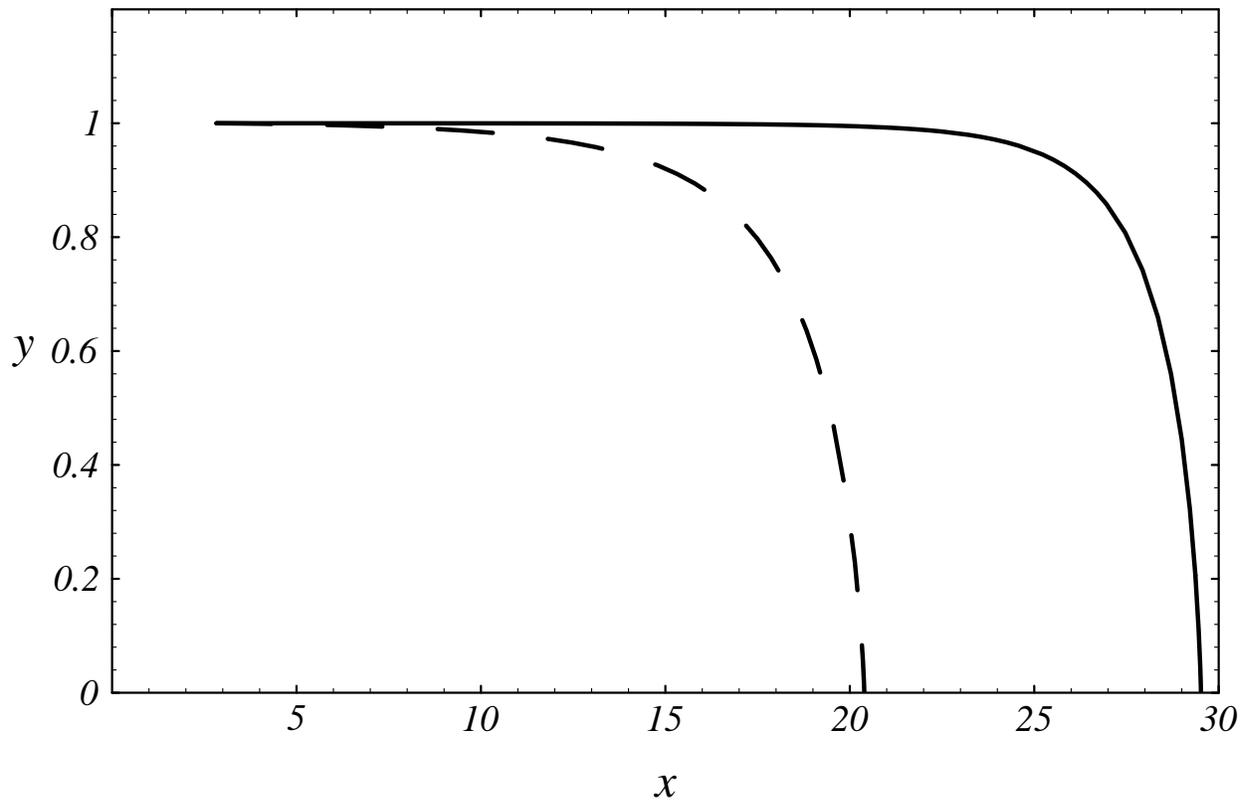
Fig.~2. Classical trajectory in the space of fields $(x,y)$
defined in Eq.~(\ref{new}). The solid line corresponds to
parameters $(a_1 = 10^{-3},  a_2 = 10^{-2})$, while
the dashed line corresponds to $(a_1 = 10^{-2},  a_2 =
10^{-2})$.

\

\noindent
Fig.~3a. The solid line shows the spectrum of density perturbations at
the end of inflation, ${\cal P}_{\zeta_e}(k)$, as a function of the
number $N_*$ of $e$-folds before the end of inflation, when the
corresponding scale left the horizon, for parameters $(a_1 =
10^{-3}, a_2 = 10^{-2})$. The dashed line corresponds to ${\cal
P}_{Q_1}(k)$ and the dotted line to ${\cal P}_{Q_3}(k)$.

\

\noindent
Fig.~3b. Same as in Fig.~2a, but for parameters
$(a_1 = 10^{-2},  a_2 = 10^{-2})$.

\

\noindent
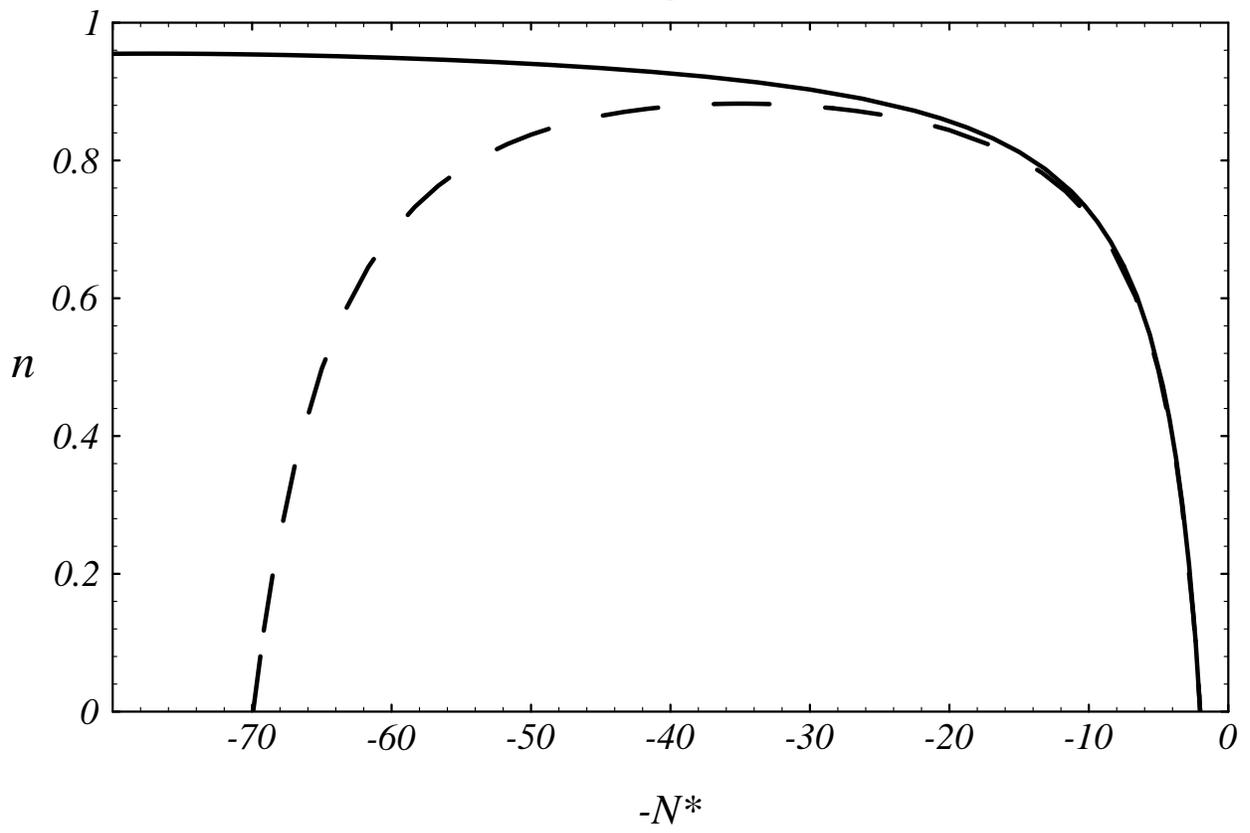
Fig.~4. The tilt $n(k)$ of the spectrum of density perturbations,
as a function of the number $N_*$ of $e$-folds before the
end of inflation, when the corresponding scale left the
horizon. The solid line corresponds to parameters
$(a_1 = 10^{-3},  a_2 = 10^{-2})$, while the dashed
line corresponds to $(a_1 = 10^{-2},  a_2 = 10^{-2})$.

\

\noindent
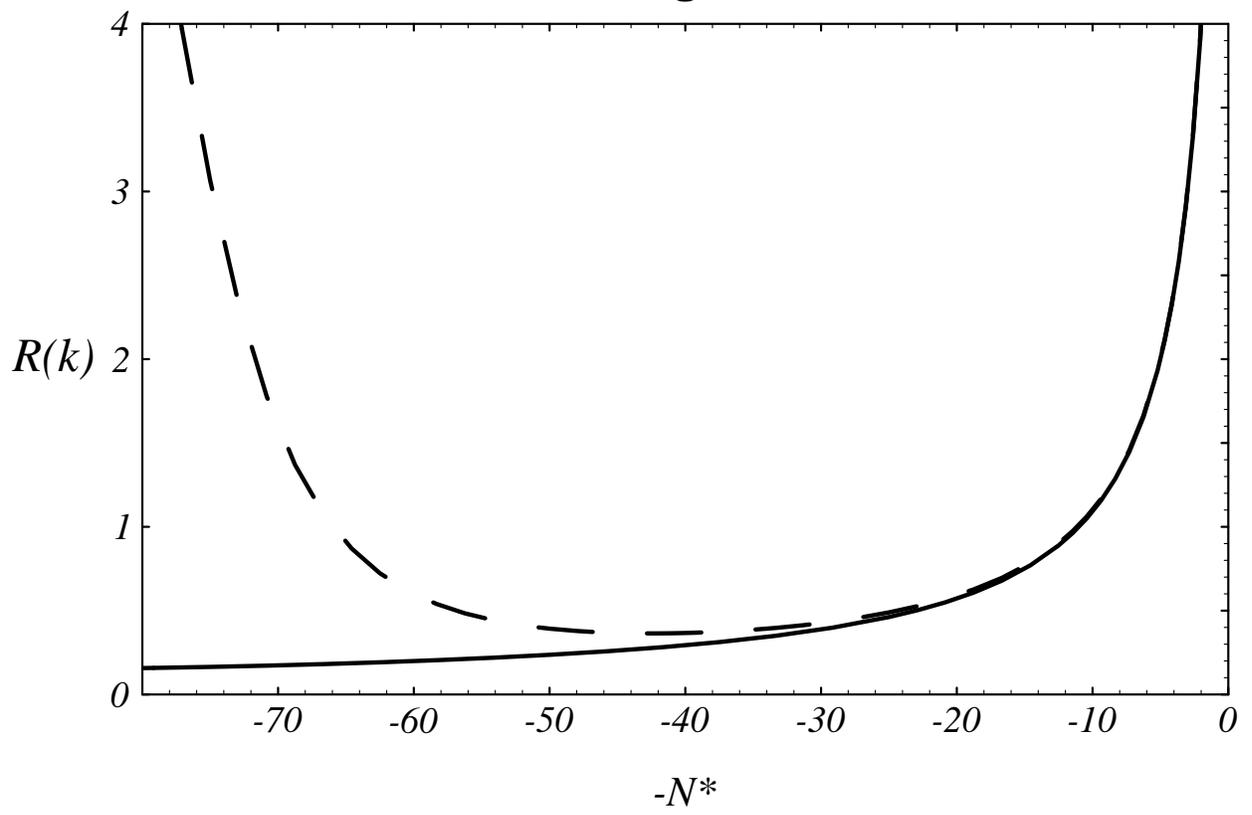
Fig.~5. The ratio $R(k)$ of tensor (gravitational waves) to scalar
(density) perturbations, as a function of the number $N_*$ of
$e$-folds before the end of inflation, when the corresponding
scale left the horizon. The solid line corresponds to parameters
$(a_1 = 10^{-3},  a_2 = 10^{-2})$, while the dashed
line corresponds to $(a_1 = 10^{-2},  a_2 = 10^{-2})$.

\

\noindent
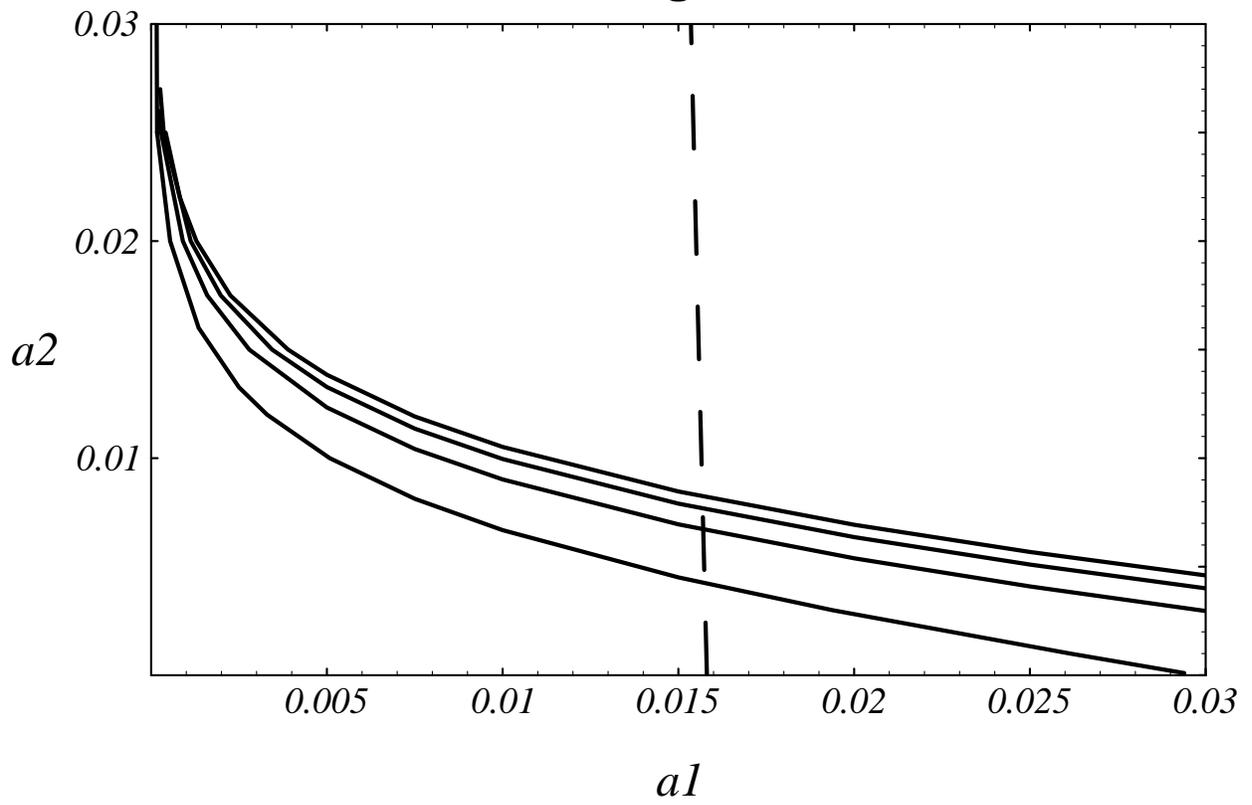
Fig.~6. Contour plot for the tilt $n_s$ of the spectrum of
density perturbations in the $(a_1,a_2)$ parameter-space.
The region below the curves is the allowed
region for $n > 0.6, 0.7, 0.8$ and $0.9$, from top to bottom.
$(a_1,a_2)$ corresponds to $(\alpha,\alpha')$ at the end of
inflation. For comparison, the dashed line corresponds to the
post-Newtonian bounds on $(\alpha,\alpha')$ in Eq.~(\ref{PNB}).

\

\noindent
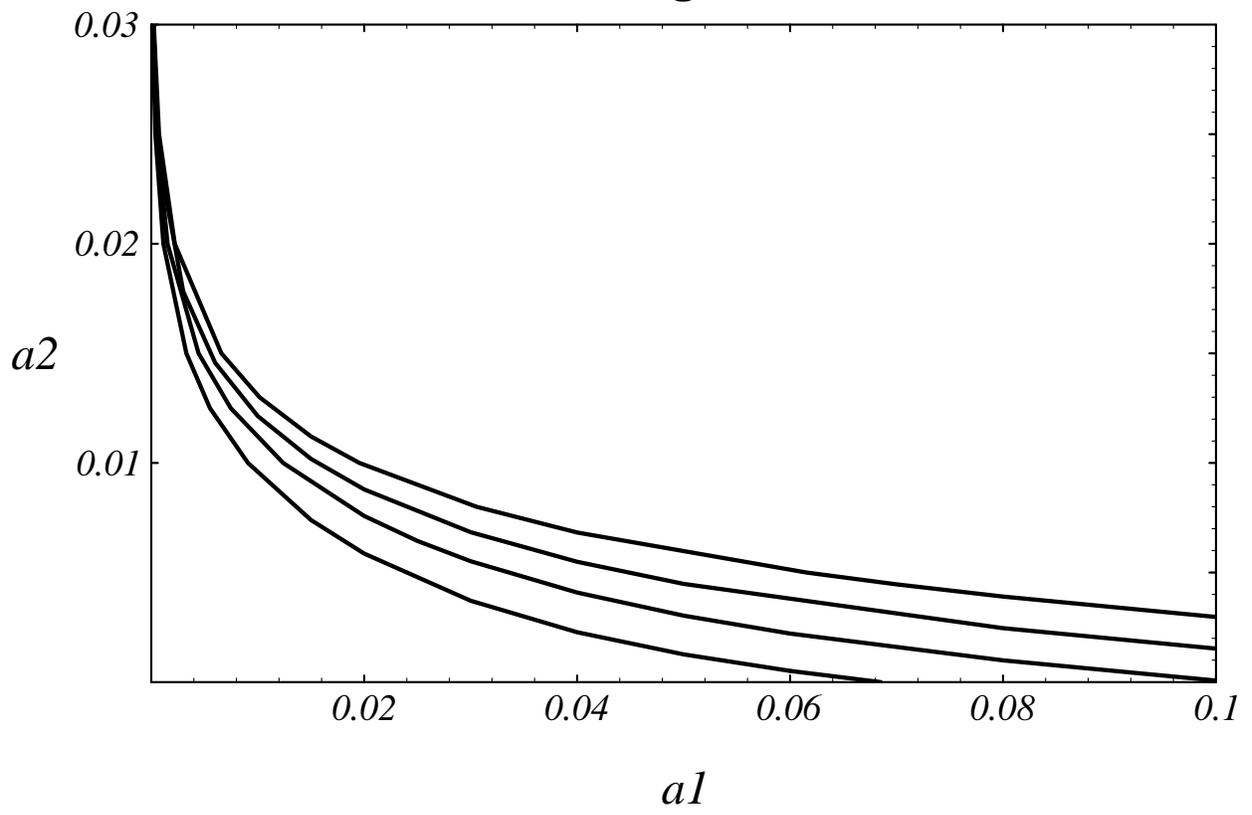
Fig.~7. Contour plot for the ratio $R$ of tensor to scalar
perturbations in the $(a_1,a_2)$ parameter-space.  The region below
the curves is the allowed region for $R < 4, 2, 1$ and $0.5$, from top
to bottom.  $(a_1,a_2)$ corresponds to $(\alpha,\alpha')$ at the end
of inflation.

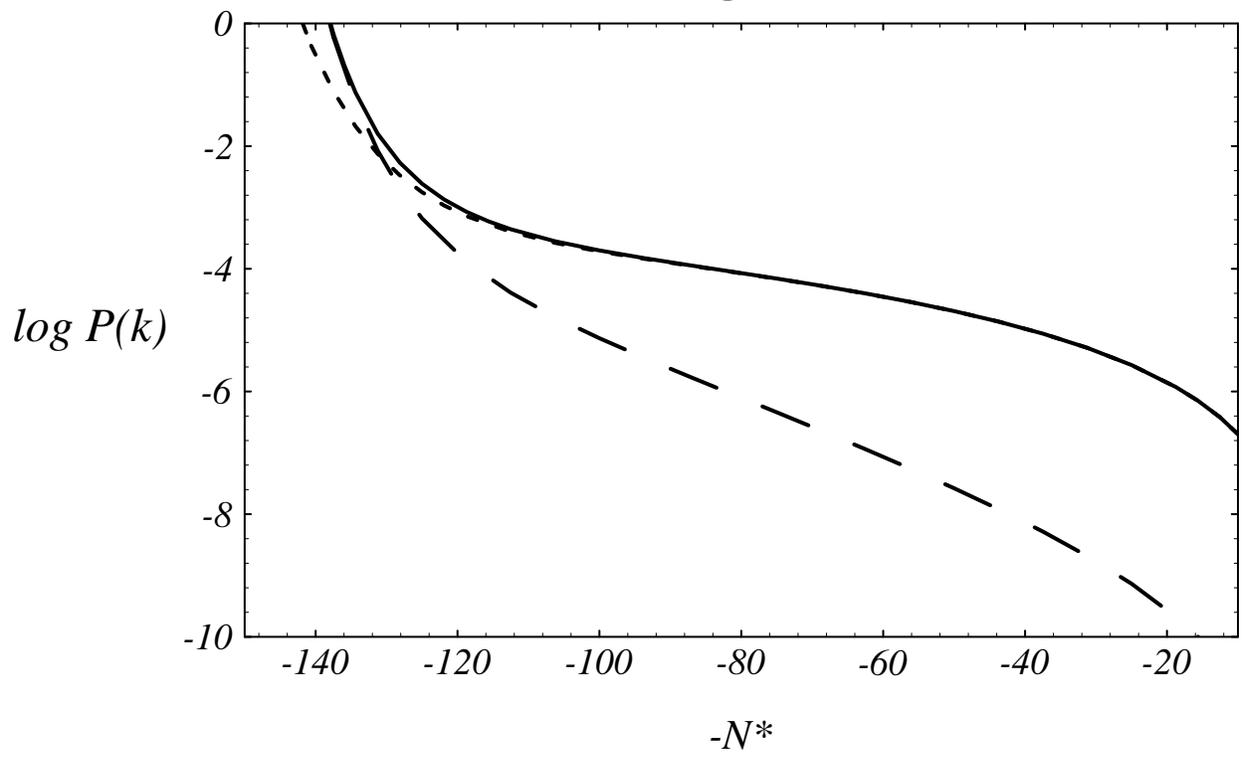
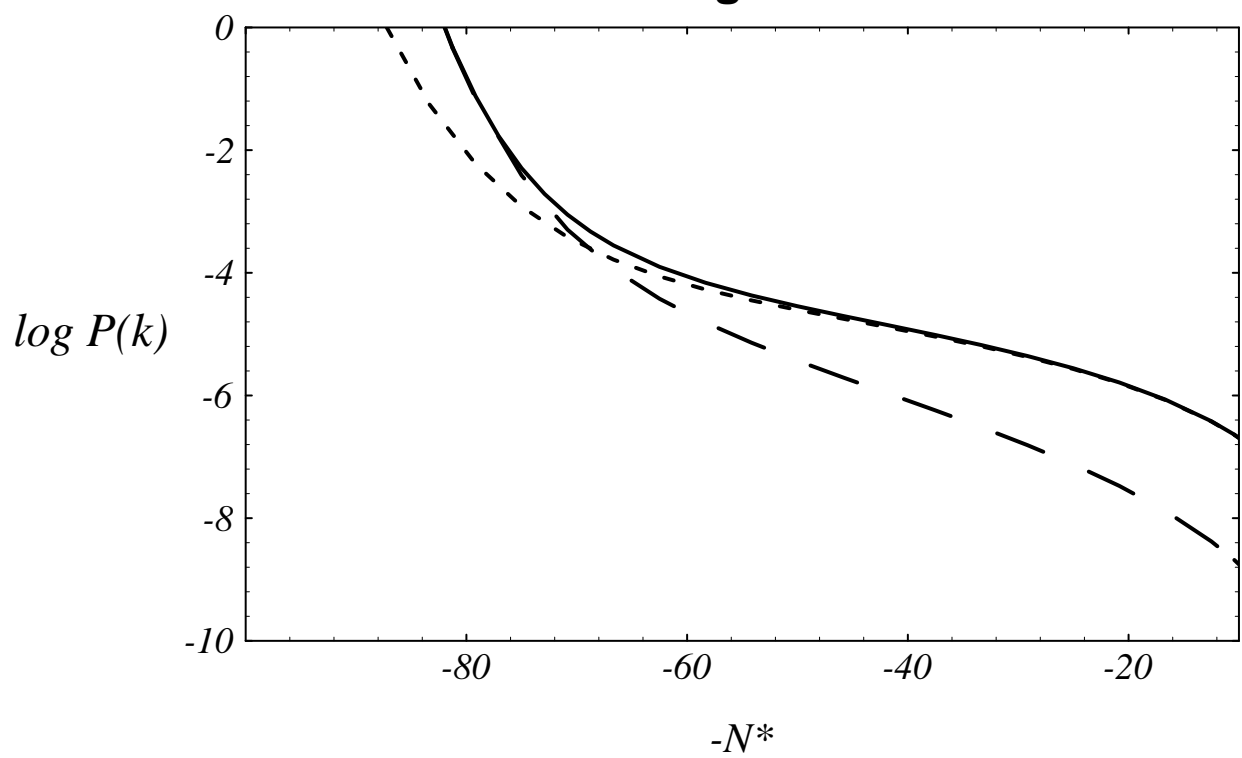
\end{document}